\documentclass[pra,superscriptaddress,twocolumn,nofootinbib]{revtex4}

\usepackage{amsmath,amssymb,graphicx}
\usepackage{color}

\def\rootfig{./figures/}

\begin{document}

\date{\today}

\title{\textbf{Stabilization of ring dark solitons in
Bose-Einstein condensates}}

\author{Wenlong Wang}
\email{wenlong@physics.umass.edu}
\affiliation{Department of Physics, University of Massachusetts,
Amherst, Massachusetts 01003, USA}
\author{P.G. Kevrekidis}
\email{kevrekid@math.umass.edu}
\affiliation{Department of Mathematics and Statistics, University of Massachusetts,
Amherst, Massachusetts 01003-4515, USA}
\affiliation{Center for Nonlinear Studies and Theoretical Division, Los Alamos
National Laboratory, Los Alamos, NM 87544, USA}
\author{R. Carretero-Gonz{\'a}lez}
\affiliation{Nonlinear Dynamical Systems Group\footnote{URL: {\tt http://nlds.sdsu.edu/}},
Department of Mathematics and Statistics,
and Computational Sciences Research Center\footnote{URL: {\tt http://www.csrc.sdsu.edu/}},
San Diego State University, San Diego, California 92182-7720, USA}
\author{D. J. Frantzeskakis}
\affiliation{Department of Physics, University of Athens,
Panepistimiopolis, Zografos, Athens 15784, Greece}
\author{Tasso J. Kaper}
\affiliation{Department of Mathematics and Center for BioDynamics,
Boston University, Boston MA 02215, USA}
\author{Manjun Ma}
\email{mmj@cjlu.edu.cn}
\affiliation{Department of Mathematics, School of Science,
Zhejiang Sci-Tech University, Hangzhou, Zhejiang, 310018, China}

\begin{abstract}
Earlier work has shown that ring dark solitons in two-dimensional
Bose-Einstein condensates are generically unstable. In this work,
we propose a way of stabilizing the ring dark soliton via a
radial Gaussian external potential. We investigate the
existence and stability of the ring dark soliton upon variations of
the chemical potential and also of the strength of the radial potential.
Numerical results show that the ring dark soliton can be stabilized in
a suitable interval of external potential strengths and chemical potentials.
We also explore different proposed particle pictures considering the ring as a
moving particle and find, where appropriate, results in very good qualitative and
also reasonable quantitative agreement with the numerical findings.
\end{abstract}


\maketitle

\section{Introduction}

Over the past few years, there has been an intense research
interest, not only theoretically, but also experimentally,
in the physics of atomic Bose-Einstein condensates (BECs)~\cite{book1,book2}, 
and particularly in the study of nonlinear 
waves~\cite{emergent}. 
Bright~\cite{expb1,expb2,expb3},
dark~\cite{djf} and gap~\cite{gap} matter-wave solitons, as well as
vortices~\cite{emergent,fetter1,fetter2}, 
solitonic vortices and vortex rings~\cite{komineas_rev} are only some among
the many structures studied (including more exotic ones
such as Skyrmions~\cite{bigelow} or Dirac monopoles~\cite{david}).

One of the most prototypical excitations that have been intensely studied
in experiments are dark
solitons~\cite{djf}. While the early experiments
in this theme were significantly limited by dynamical instabilities
and thermal effects~\cite{han1,nist,dutton,han2,nsbec}, more recent efforts
have been significantly more successful in generating and exploring
these structures. By now, the substantial control of the generation,
and dynamical interactions of such structures has led to a wide
range of experimental works monitoring their evolution in different
settings~\cite{engels,Becker:Nature:2008,hambcol,kip,andreas,jeffs}.

The instability of dark solitons in higher dimensions
(towards bending~\cite{nist} and eventual snaking towards vortices/vortex
rings~\cite{nsbec,beckerus})
has been one of the key reasons for the inability to study such states
in higher dimensions. Although external stabilization mechanisms, e.g.,
utilizing a blue-detuned laser beam~\cite{manjun}, have been proposed,
importantly also variants of such dark solitons have been explored
in higher dimensions in the form of {\it ring dark solitons} (RDSs). These
efforts were at least in part motivated by works in nonlinear optics,
where 
they initially were introduced in Ref.~\cite{KYR}, and 
studied in detail, both theoretically (in conservative~\cite{thring,djfbam,ektoras}
and ---more recently--- in dissipative \cite{dum} settings)
and experimentally~\cite{expring1}. In turn, RDSs in BECs 
were originally proposed in Ref.~\cite{rds2003} and their dynamics
was analyzed by means of the perturbation theory of dark matter-wave
solitons \cite{djf}. In other works, RDSs were studied by different approaches,
e.g.,
in a radial box~\cite{ldc1}, by
using a quasi-particle approach \cite{kamch}, or by considering them as
exact solutions in certain versions of the Gross-Pitaevskii equation (GPE) \cite{toik1}.
Proposals for the creation of RDS, e.g., by means of BEC self-interference \cite{ch1} or by
employing the phase-imprinting method \cite{song}, as well as
generalizations of such radial states (including multi-nodal ones)
\cite{ldc1,herring} have also been considered. 
Moreover, generalizations of RDSs
were studied in multi-component settings, in the form of
dark-bright ring solitons~\cite{stockhofe}
(emulating the intensely studied context of multi-component one-dimensional (1D)
dark-bright solitons~\cite{buschanglin,sengdb,peter1,peterprl}), or
in the form of vector RDS in spinor $F=1$ BECs \cite{song}. Importantly,
structures of the form of
radially symmetric dark solitons, closely connected to RDSs
exist also in three-dimensions
with a
spherical rather than cylindrical symmetry
(so-called ``spherical shell solitons'' \cite{ldc1}).

Nevertheless, in none of these contexts (either one- or multi-component),
was it possible to achieve complete stabilization of the RDSs. In particular,
stabilization mechanisms that have been proposed, e.g., by ``filling'' the
RDS by a bright soliton component \cite{stockhofe} or by employing the nonlinearity,
management (alias ``Feshbach resonance management'' \cite{FRM}) technique \cite{ch2},
were only able to prolong the RDSs' life time.
In fact, it was illustrated that the instabilities of the ring-shaped solitons were
connected, bifurcation-wise, to the existence of vortex ``multi-poles'',
such as vortex squares (which are generically stable in evolutionary
dynamics), vortex hexagons, octagons, decagons etc.; all of these
states are progressively more unstable. This picture has been
corroborated by detailed numerical computations in Ref.~\cite{middelphysd}.
It is our aim in this work to revisit the RDSs and
their destabilization mechanisms and, indeed, to propose a technique
for their complete dynamical stabilization. Our technique is
reminiscent of that of Ref.~\cite{manjun} in that we introduce a potential induced by a
{\it radial} blue-detuned laser beam. Radial potentials of a similar
form have been intensely used in recent experiments, e.g., by the
groups of~\cite{gretchen} and~\cite{boshier}
and are hence accessible to state-of-the-art
experimental settings.

Our presentation of this effort to stabilize the RDS in the form of
a dynamically robust state of quasi-two-dimensional BECs can be
summarized as follows: we introduce, in Sec.~\ref{setup},
the mathematical model and
our specific proposal towards a potential stabilization of the RDS. We also
incorporate in this Section theoretical attempts to explore the
coherent dynamics of the ring soliton 
by means of a particle model.
Our numerical results are presented in Sec.~\ref{results}, initially
revisiting (for reasons of completeness and to facilitate the exposition)
the case without the external radial barrier potential and
subsequently incorporating it in the picture. Finally, our concluding
remarks are presented in Sec.~\ref{conclusion}, and a number of
important open future directions is also highlighted.

\section{Model and mathematical set-up}
\label{setup}

\subsection{The Gross-Pitaevskii equation}

In the framework of lowest-order mean-field theory, and for sufficiently low-temperatures,
the dynamics of a quasi-2D (pancake-shaped) BEC confined in
a time-independent trap $V(r)$ is described by the following dimensionless GPE \cite{emergent}:
\begin{equation}
i \psi_t=-\frac{1}{2} \nabla^2 \psi+V(r) \psi +| \psi |^2 \psi-\mu \psi,
\label{GPE}
\end{equation}
where $\psi(x,y,t)$ is the macroscopic wavefunction of the BEC,
$\mu$ is the chemical potential, and $V(r)$ (with $r=\sqrt{x^2+y^2})$ is the external potential.
The latter, is assumed to be a combination of a standard parabolic
(e.g., magnetic)
trap, $V_{\rm MT}(r)$, and a localized radial ``perturbation potential'', $V_{\rm pert}(r)$, namely:
\begin{equation}
V(r) = V_{\rm MT}(r) + V_{\rm pert}(r) = \frac{1}{2} \Omega^2 r^2 + V_{\rm pert}(r),
\label{eq:VMT+Vpert}
\end{equation}
with $\Omega$ being the effective strength of the magnetic trap.
For the numerical results in this work we chose a nominal value
of $\Omega=1$ unless stated otherwise.
As will be evident from the scaling of our findings below, the particular 
value of $\Omega$ will not play a crucial role in our conclusions. 

The GPE in the Thomas-Fermi (TF) limit of large $\mu$
has a well known ground state $\psi_{\rm{TF}}=\sqrt{\max(\mu-V,0)}$.
The other interesting limit is the linear one where
the self-interaction term can effectively be ignored. In this limit, the GPE reduces
to the 2D 
harmonic oscillator problem. Both limits are
particularly useful for our considerations: the former
enables the consideration
of the ring-shaped soliton as an effective particle, 
the latter enables the construction of the ring as an exact solution in the linear limit,
which 
is continued in the nonlinear regime.

Here, we will focus on the single RDS which, in the linear limit,
can be viewed
as a superposition of the $|2,0\rangle$ and $|0,2\rangle$ 
quantum harmonic oscillator states, namely:
\begin{equation}
|\psi_{\rm{RDS}}\rangle_{\rm{linear}}=\frac{|2,0\rangle+|0,2\rangle}{\sqrt{2}}
\propto (r^2-1) e^{-\Omega r^2/2}.
\end{equation}
This linear state, which exists for $\mu > 3 \Omega$ (i.e., beyond the
corresponding linear limit of the above degenerate $n+m=2$ states,
where $n$ and $m$ are the respective indices along the $x$- and
$y$-directions, characterizing the quantum harmonic oscillator
sate $|n,m\rangle$),
can be continued to higher chemical potentials. However, the RDS
is known to be inherently unstable for all values of $\mu$ beyond
the linear limit~\cite{ldc1,herring,toddric}. This instability
breaks the original radially symmetric state into vortex
multi-poles, as originally shown in Ref.~\cite{rds2003} and
subsequently examined from a bifurcation perspective in Ref.~\cite{middelphysd}.
Our scope is to provide a systematic understanding of the RDS 
instability modes and how to suppress them, so as to potentially enable its
experimental realization. Similar considerations in the context
of exciton-polariton condensates (where a larger range of tunable
parameters exists due to the open nature of the system and the presence
of gain and loss) have led both to the theoretical analysis~\cite{rodr}
and to the experimental observation~\cite{sanv} of {\em stable} RDSs. 

Following the motivation of the earlier work of Ref.~\cite{manjun} on
planar dark solitons, in conjunction with the recent experimental
developments in the context of radial~\cite{gretchen} and more
broadly, in principle arbitrary, so-called painted~\cite{boshier}
potentials, we propose the following form for $V_{\rm pert}(r)$:
\begin{equation}
V_{\rm{pert}}(r)=Ae^{-(r-r_c)^2/(2\sigma^2)},
\end{equation}
where $r_c$, $A$ and $\sigma$ represent, respectively, the radius, the amplitude and the width
of this ring-shaped potential.

Since RDSs feature radial symmetry, we first express Eq.~(\ref{GPE})
in the form:
\begin{equation}
i \psi_t=-\frac{1}{2}\left(\frac{d^2}{dr^2}+\frac{1}{r}\frac{d}{dr}\right)\psi+V(r) \psi +| \psi |^2 \psi-\mu \psi.
\label{GPE1D}
\end{equation}
We also assume that a stationary RDS state, $\psi=\psi(r,t)$, governed by the effectively 1D model (\ref{GPE1D}),
is characterized by a radius $r_c$. In other words,
we will hereafter opt to locate the perturbation potential at the fixed equilibrium position of
the RDS.
For our analysis, the control parameters will 
be the strength $A$ of the perturbation potential and the nonlinearity strength 
(characterized by the chemical potential $\mu$); as concerns the width $\sigma$ of $V_{\rm{pert}}(r)$,
it will be fixed (unless otherwise stated) to the value $\sigma=1$, which is of the order of the soliton width
---i.e., of the healing length.

Below, we proceed with the study of the effect of the perturbation potential
on the existence and stability of the RDS.
Stability will be studied from both the spectral perspective,
through a Bogolyubov-de Gennes (BdG) 
analysis, and from a dynamical time evolution perspective.
The latter, will involve direct numerical integration of Eq.~(\ref{GPE}), whereby 
a (potentially perturbed) RDS is initialized and its evolution is monitored
at later times.
On the other hand, BdG analysis for a stationary RDS, $\psi_0(r)$, will involve the study of the eigenvalue
problem stemming from the linearization of Eq.~(\ref{GPE}), upon using the perturbation ansatz:
\begin{eqnarray}
\psi(x,y,t)=\psi_0(r) + \delta \left(u(x,y) e^{\lambda t}
+ \upsilon^{\ast}(x,y) e^{\lambda^{\ast} t} \right),~~
\label{bdg}
\end{eqnarray}
where $[\lambda,(u,\upsilon)^T]$ is the eigenvalue-eigenvector pair, $\delta$
is a formal small parameter, and the asterisk
denotes complex conjugation.
Then, the existence of eigenvalues with non-vanishing real
part signals the presence of dynamical instabilities. These
come in two possible forms: (a) genuinely real eigenvalue pairs,
which are associated with an exponential instability; and
(b) complex eigenvalue quartets that denote an oscillatory instability,
where growth is coupled with oscillation. The above symmetry of
the eigenvalue pairs (i.e., the fact that they only arise in
pairs or quartets) stems from the Hamiltonian nature of the problem.

\subsection{The particle picture for the ring dark soliton}
\label{sec:sub:part}

A natural way 
to obtain a reduced dynamical description of the RDS is to adopt
a particle picture and use a variational approximation
discussed in detail in Ref.~\citep{oberthaler}.
According to this approach, in the TF limit (i.e., for sufficiently large chemical potential),
the RDS state can be approximated by
a product of the
TF ground state, $\psi_{\rm{TF}}=\sqrt{\max(\mu-V,0)}$,
and a (potentially traveling)
dark soliton of radial symmetry, 
of the form:
\begin{equation}
\psi_{\rm DS}(r,t)=  b(t) \tanh[\sqrt{\mu}\,b(t)(r-r_c(t))]+ia(t),
\label{pe}
\end{equation}
where $b$ and $a$ (with $a^2+b^2=1$) set, respectively, the depth and velocity of the soliton, while
$r_c$ is the RDS radius. Then, the Euler-Lagrange
equations for the two independent
effective variational parameters $r_c$ and $a$, stemming from the averaged renormalized Lagrangian of the system,
take the following form~\cite{oberthaler}:
\begin{eqnarray}
\label{a}
\dot{a}&=&-\frac{b^2}{\sqrt{\mu}}
\left\{\left(\frac{V'}{2}-\frac{\mu}{3r_c}\right)+\frac{VV'}{3\mu} \right. \nonumber \\
&&\left. +V'\left[\frac{V^2}{3\mu^2}+\frac{1}{4}\left(\frac{2}{3}-\frac{\pi^2}{9}\right)\frac{V'^2}{\mu^2}\right]\right\}, \\
\label{b}
\dot{r}_c&=&\sqrt{\mu}\left[a\left(1-\frac{V}{2\mu}\right) \right. \nonumber \\
&&
\left.-\frac{a}{4b^2}\left(\frac{5}{3}-\frac{\pi^2}{9}\right)\frac{V'^2}{\mu}\left(1-\frac{2V}{\mu}\right)\right].
\end{eqnarray}
The above system suggests the existence of stationary RDSs,
due to the interplay (to the leading-order
approximation in $\Omega$) of an effective attractive trapping potential and an
effective curvature-induced repulsive logarithmic potential ---see first and second terms in the right-hand side of
Eq.~(\ref{a}), respectively. A more systematic analysis, that takes into regard higher-order terms in $\Omega$,
shows that the critical radius for which a stationary ring exists is given
by~\cite{oberthaler}
\begin{equation}
r_c=\frac{\sqrt{0.5616 \mu}}{\Omega}.
\label{appr1}
\end{equation}
Notice that, according to the discussion of Ref.~\cite{oberthaler}
and in accordance with the computational analysis
presented below (see Sec.~\ref{sec:results:particle}),
the numerical results strongly suggest an
asymptotic critical radius  $r_c=\sqrt{\mu/2}/\Omega$
(see also the discussion in Refs.~\cite{oberthaler,rds2003,kamch}).

This discrepancy suggests the consideration of
alternative ways of determining the stationary RDS' radius.
Here, for reasons of completeness, we will present such an alternative approach,
based on the earlier work of Ref.~\cite{kaper} for a different system (namely, ring-like steady state
solutions of coupled reaction-diffusion equations).
More specifically, our starting point will be the steady state problem
associated with Eq.~(\ref{GPE1D}), where we will ``lump'' the potential
terms as $V(r)=V_{\rm MT} + V_{\rm pert}(r)$. Using the ansatz
$\psi(r)=\psi_{\rm TF}(r) q(r)$, we obtain the steady state problem:
\begin{eqnarray}
\frac{1}{2} q'' + \mu q (1-q^2) = P(r),
\label{neweq}
\end{eqnarray}
where $$P(r)= V q (1-q^2) -\frac{q'}{2 r}
-\frac{\psi_{\rm TF}''}{2 \psi_{\rm TF}} q
- \frac{\psi_{\rm TF}'}{\psi_{\rm TF}} q'
- \frac{1}{r} \frac{\psi_{\rm TF}'}{2 \psi_{\rm TF}} q,$$
and primes denote derivatives with respect to $r$.
Then, seeking a stationary RDS solution in the form of $q(r)=\tanh(\sqrt{\mu}
(r-r_c))$
and multiplying both sides by
$q'$ in Eq.~(\ref{neweq}), we 
find that the left-hand side is simply $dH/dr$, where $H$ is the effective Hamiltonian
$H=q'^2/4 - \mu (1-q^2)^2/4$.
Hence, upon integrating in $r$ from $-\infty$ to $\infty$, bearing in mind that the error
between $r=0$ and $r \rightarrow -\infty$ is exponentially small,
we obtain the explicit solvability, Melnikov-type, condition~\cite{GH}:
\begin{eqnarray}
\int_{-\infty}^{\infty} P(r) q'(r) dr=0.
\label{neweq2}
\end{eqnarray}
Upon evaluating the integrals of all five terms associated with
$P(r)$ within Eq.~(\ref{neweq2}), we should obtain
an algebraic equation
for the equilibrium position of the RDS.
Indeed, evaluating the first potential term (for $A=0$),
through a series of rescalings and integrations by parts, leads to
$\Omega^2 r_c/(3\sqrt{ \mu})$. In turn, the second term yields
$-2 \sqrt{\mu}/(3 r_c)$ and the fourth term yields
$2 \Omega^2 r_c \sqrt{\mu}/[3 (\mu-\Omega^2 r_c^2/2)]$,
while the other terms contribute at higher order.
Putting all the terms together
in the case of $A=0$ yields the prediction
\begin{eqnarray}
r_c=\frac{\sqrt{\alpha\mu}}{\Omega},
\label{neweq3}
\end{eqnarray}
where $\alpha=4-2\sqrt{3} \approx 0.5359$; this result is
more accurate than the one of Eq.~(\ref{appr1}), as will be discussed in more detail in
Sec.~\ref{sec:results:particle}.

Finally, we proceed to give a third method, based on the
analysis of Ref.~\cite{kamch}, that will prove to be the
most accurate one in connection to our computations of not only statics
but also dynamics of RDS states in the numerical section that will follow.
In the latter approach, it is argued that the equation of motion
can be derived by a local conservation law (i.e., an adiabatic
invariant) in the form of the energy of a dark soliton under the
effect of curvature and of the density variation associated with it.
More specifically, knowing that the energy of the one-dimensional dark
soliton is given by~\cite{djf} ${\cal E}_{\rm DS}=(4/3) (\mu - \dot{x_c} )^{3/2}$,
where $x_c$ is the dark soliton position,
the generalization of the relevant quantity in a two-dimensional
domain bearing density modulations reads:
\begin{equation}
{\cal E}_{\rm RDS}(r) = 2 \pi r \left[ \frac{4}{3} (\mu-V(r)-\dot{r}^2)^{3/2} \right].
\label{PPnew1}
\end{equation}
Thus, by assuming this quantity is constant, namely
${\cal E}_{\rm RDS}(r)={\cal E}_{\rm RDS}(r_c)$, where
$r_c$ is the equilibrium location of the ring,
we obtain an equation for $\dot{r}^2$.
Taking another time derivative on both sides,
we finally obtain Newtonian particle dynamics for the ring in the form:
\begin{equation}
\ddot{r}=-\frac{1}{2}\frac{\partial V}{\partial r}
+\frac{1}{3 r}\left(\frac{r_c}{r}\right)^{2/3}\left[\mu-V(r_c)\right].
\label{PPextra}
\end{equation}
When $A=0$, this equation of motion for the RDS position yields
the equilibrium
$r_c=\sqrt{{\mu}/{2}}/\Omega$, a result which, as highlighted also
above and as will be demonstrated below, is the one most consistent with
the numerical observations. This, in turn, motivates us to use the
above approach of Ref.~\cite{kamch} not only for the statics, but also
for the dynamics in the following section and additionally, not only for
the case without the radial defect of $A=0$, but also for that
bearing the radial defect i.e., for $A \neq 0$.

\begin{figure}[tb] 
\begin{center}
\includegraphics[width=8.5cm]{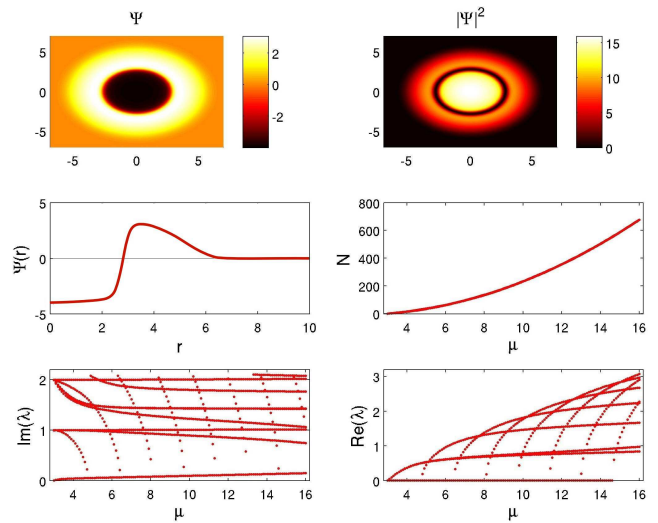}
\caption{
(Color online) Top panels: the RDS'
real-valued profile (left)
and the corresponding density plot (right) for $\mu=16$.
Middle left: a radial profile of the relevant state.
Middle right: 
number of particles $N=\int |\psi|^2dr$ as a function of chemical potential $\mu$, showing the
continuation of states from the linear limit to the nonlinear regime.
Bottom panels: the imaginary (left) and real parts (right) of the spectrum;
showcased is the generic instability of the RDS, and the emergence
of additional unstable eigenmodes thereof as $\mu$ is increased.
The value for the trap strength in this figure and all remaining
figures is $\Omega=1$ (unless stated otherwise).
}
\label{lc}
\end{center}
\end{figure}

We now proceed to test these predictions, as well as to examine the
BdG stability analysis and the dynamical evolution of the RDS, both
in the absence (initially, for comparison and guidance) and then
in the presence of the radial perturbation 
potential.

\section{Results}
\label{results}

First, we briefly summarize the numerical techniques used in this work.
Stationary states in both 1D (i.e., in a radial form) and 2D were identified
using a centered finite-difference scheme within Newton's method. The
spectrum of the stationary states (i.e., the result of the BdG analysis)
was calculated using the eigenvalue problem derived from Eq.~(\ref{bdg}).
Finally, for the dynamics of the system, we used direct integration employing
second order finite differences in space and fourth-order Runge-Kutta in time.

\subsection{Basic properties of the ring dark soliton}
Let us start by summarizing some of the basic properties of the
RDS \textit{without the perturbation potential.} A typical RDS state in the TF limit
of large chemical potential $\mu$
is shown in the top and middle left panels of Fig.~\ref{lc}; the
top right panel shows the corresponding density. As indicated in the
previous section, the RDS has a linear limit (built out of the
eigenstates of the 2D quantum harmonic oscillator). The 
continuation of such a state in the nonlinear regime is shown in the middle right panel of
Fig.~\ref{lc}. The imaginary and real parts of the spectrum of the
RDS are shown in the bottom panels of Fig.~\ref{lc}. Note that the
RDS is unstable for any value of $\mu$ beyond the linear limit. More importantly,
in line with what was also presented in Ref.~\cite{middelphysd}, as $\mu$
increases, more unstable modes keep emerging, through eigenvalue pairs
that cross through the origin. These signal pitchfork bifurcations,
to which we now turn.

\begin{figure}[tb] 
\begin{center}
\includegraphics[width=8.5cm]{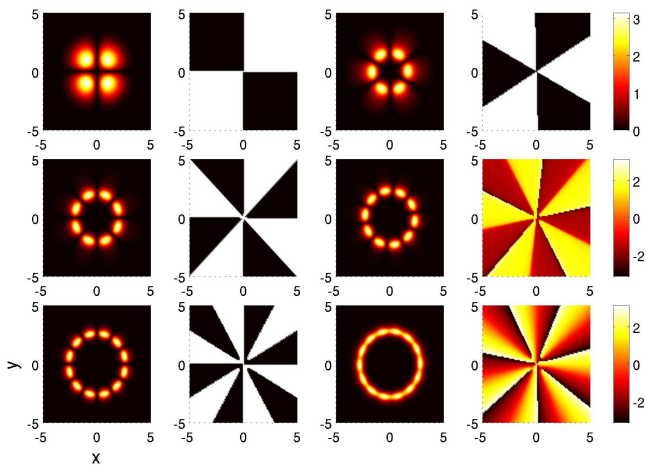}
\caption{(Color online) The most unstable modes at a few representative chemical potential
values $\mu=4$, 6, 9, 11, 14, and 16 (from left to right, top to bottom)
associated with the instability of the RDS.
Left and right subpanels correspond, respectively, to the absolute value
and phase of the modes.
}
\label{evecs}
\end{center}
\end{figure}

\begin{figure}[tb] 
\begin{center}
\includegraphics[width=8.5cm]{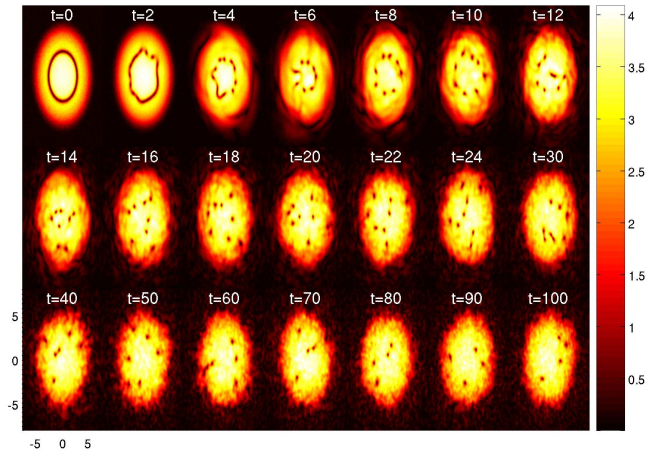}
\caption{(Color online) Dynamics ensuing from the unstable RDS for $\mu=16$.
Note that the RDS first deforms into seven
pairs of vortices (in accordance with the most unstable mode
for these parameters values; see the bottom  right panel of
Fig.~\ref{evecs}), and then eventually turns into a dynamical evolving
vortex cluster for longer times.
During evolution, some of the vortices are ``absorbed'' by the
BEC periphery and the system is eventually left with four
interacting vortices.}
\label{RK4A0}
\end{center}
\end{figure}

Studies of RDS in atomic BECs have illustrated their dynamical
breakup into vortex-antivortex pairs (see, e.g., Refs.~\cite{rds2003,herring}).
To complement this picture, we now discuss the most unstable modes of the
BdG analysis. Some representative eigenmodes at
$\mu=4,$ 6, 9, 11, 14 and 16 are shown in Fig.~\ref{evecs}.
It is interesting to observe that the identified
modes indicate a clear connection to an increasing number of pairs of vortices.
The first unstable mode appears to be connected to two-pairs,
i.e., to a vortex quadrupole. Indeed, the vortex quadrupole exists
as a state~\cite{mottonen} for any value of $\mu$ beyond the
linear limit of $\mu=3 \Omega$, being constructed as:
\begin{equation}
|\psi_{\rm{Q}}\rangle_{\rm{linear}}=\frac{|2,0\rangle + i |0,2\rangle}{\sqrt{2}}.
\end{equation}
Subsequent destabilization modes reveal a three-fold symmetry (leading
to the bifurcation of vortex hexagons~\cite{middelphysd}), a four-fold
symmetry (leading to vortex octagons), then a five-fold (decagons),
a six-fold (dodecagons), and so on. These different eigenvectors
are clearly illustrated in Fig.~\ref{evecs} and the existence and
stability of the corresponding emerging (from the pitchfork bifurcation)
vortex $n$-gon cluster states is discussed in Ref.~\cite{annab}.

A dynamical study of the states shows that the evolution initially results
in vortex pairs, in agreement with Fig.~\ref{evecs}. However, gradually
some vortices may move out of the BEC and get lost in the background,
leaving behind a complex, interacting cloud of vortices, as shown for
$\mu=16$ in Fig.~\ref{RK4A0}. The resulting
interaction dynamics between vortices in the cluster, and the associated
transfer of energies between different scales, may represent a very
interesting setting for exploring turbulence phenomena and associated
cascades in line, e.g., with recent experimental efforts of Ref.~\cite{bpa_turb}.

\begin{figure}[tb] 
\begin{center}
\includegraphics[width=8.5cm]{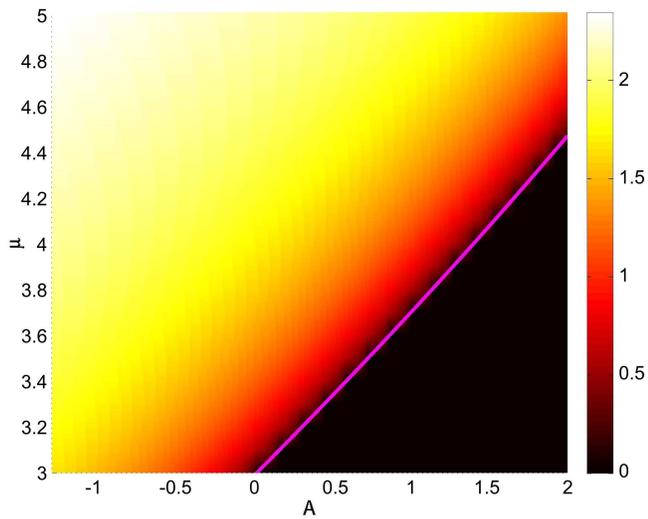}
\caption{(Color online) Max$(|\Psi|)$ as a function of $(A,\mu)$.
Note that $N$ decreases as $A$ increases when holding $\mu$ fixed until some critical set of values of $A$
(depicted by the purple line)
beyond which the RDS will cease to exist. In the linear limit
$\mu=3 \Omega$, even a very small positive perturbation of $A$ will destroy the RDS state.}
\label{surface2}
\end{center}
\end{figure}

\begin{figure}[thb] 
\begin{center}
\includegraphics[width=8.5cm]{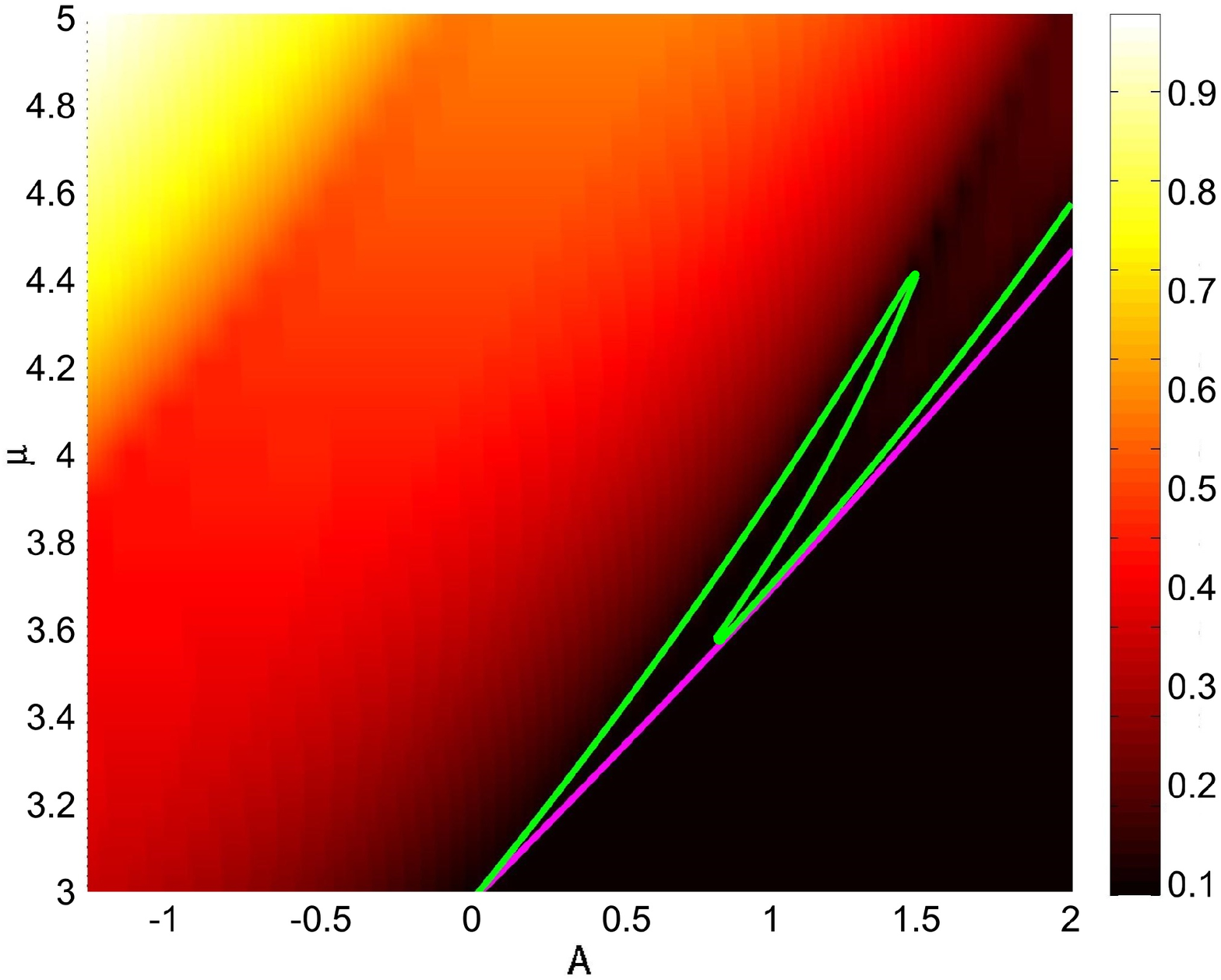}
\caption{(Color online) Instability growth rate max(Re($\lambda$)) as a function of $(A,\mu)$.
The area between the left (green)
and right (purple) curves 
corresponds to
the region where the RDS exists and with vanishing max(Re($\lambda$)),
i.e. the RDS is completely stable.
The rightmost purple line is also the boundary of the critical values of
$A$ beyond which no RDS solution exists.
}
\label{surface3}
\end{center}
\end{figure}

\subsection{Adding the perturbation potential}

Having analyzed the unperturbed case, we now examine the
case with the radial Gaussian potential. 
The existence of the RDS structure in the latter
case can be captured as a function of $(A,\mu)$
---see Fig.~\ref{surface2}.
We used max$(|\Psi|)$ (i.e., the max root density) as a diagnostic
instead of $N$ for practical visualization purposes, in this case.
We can see that for a fixed value of $\mu$, the density decreases as $A$ increases
(a natural feature, given the repulsive nature of the perturbation potential)
until a critical value of $A$ ---shown as a purple line--- is reached, beyond which
the RDS will cease to exist. In the linear limit of
$\mu=3 \Omega$, even a very small positive perturbation of $A$
will destroy the RDS state. The monotonic dependence
of $\mu$ on the critical $A$ appears to be approximately linear.

We proceed now with the central theme of this study, which
is the dynamical stabilization of the RDS. 
To characterize the stability of the RDS in the $(A,\mu)$ plane, in Fig.~\ref{surface3}
we show a plot of the max(Re($\lambda$)) as a function of $(A,\mu)$. 
The right most purple line, as before, 
depicts the critical values of $A$ beyond which no RDS solution exists. The region enclosed between the green and purple
lines corresponds to the regimes where 
RDS exists {\it with vanishing max(Re($\lambda$))}, i.e., the RDS is
completely stabilized by the presence of the external Gaussian
ring perturbation potential. One interesting feature is that the relevant
stability landscape is rather complex with potential sequences of
destabilization and restabilization for values of $\mu \geq 3.6$ (we will
return to this point below).
However, the principal conclusion obtained from Fig.~\ref{surface3} is that the
RDS is generically subject to full dynamical stabilization
for any value of the chemical potential and for suitable intervals of the
perturbation potential strength $A$ in the vicinity of the linear limit.
The feature that the stabilization is enabled near the linear limit is
rather natural to expect also on the basis of our earlier results
for $A=0$ in Fig.~\ref{lc}. Given that the RDS is progressively more
and more unstable (with a higher number of destabilizing modes) as
$\mu$ increases suggests that the perturbation potential may be unable to suppress
this multitude of unstable modes, especially far from the linear
limit.

\begin{figure}[tb] 
\begin{center}
\includegraphics[width=8.5cm]{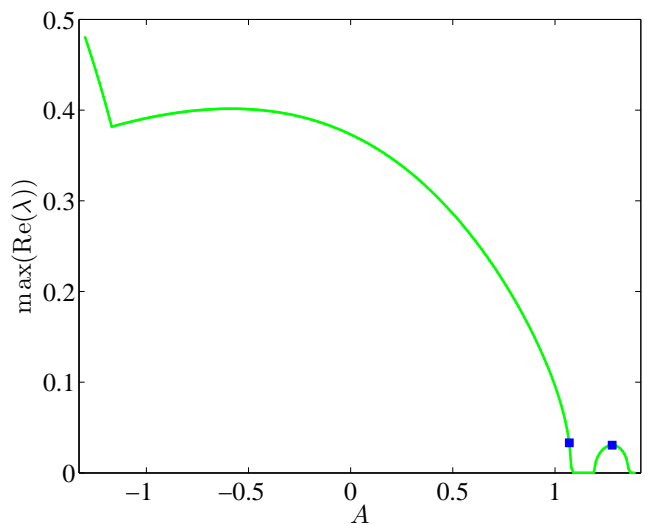}
\caption{(Color online) Cross section of the instability growth rate max(Re($\lambda$)) at $\mu=4$. The right most point of the curve
corresponds to the critical value of $A$ beyond which no RDS solution exists. The two blue squares are two points in two different instability regimes but with similar instability rates whose full spectrum is shown in Fig.~\ref{css1}.
}
\label{css}
\end{center}
\end{figure}

\begin{figure}[tb] 
\begin{center}
\includegraphics[width=8.5cm]{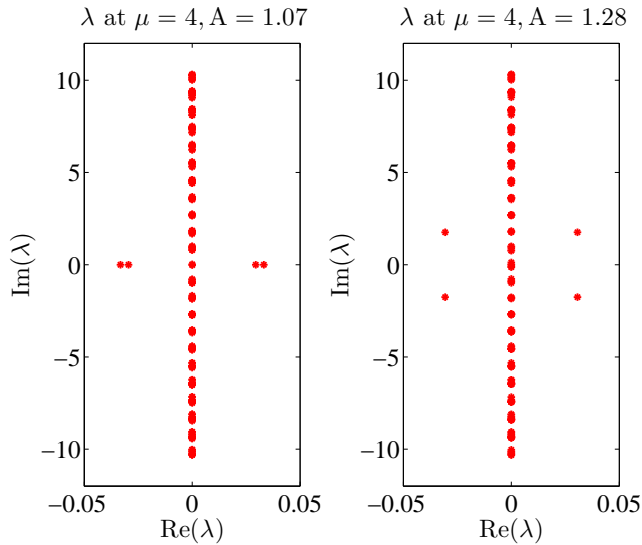}
\caption{(Color online) Full stability spectrum corresponding to the two blue
squares
in the two different instability regimes in Fig.~\ref{css} for $\mu=4$.
Left and right panels correspond to $A=1.07$
and $A=1.28$, respectively.
Note that the two regimes do not share the same nature of instability.
The large amplitude case (left) has the instability on the real axis
(i.e., exponential instability) while
the small amplitude case (right) has the instability in the form of
a complex quartet (oscillatory instability).}
\label{css1}
\end{center}
\end{figure}

To gain further insight on this stability plane, let us
now study a typical cross section of Fig.~\ref{surface3} at $\mu=4$.
The cross section is shown in Fig.~\ref{css}. A detailed study of the full
spectrum shows the existence
of two intervals of instability which are not of the same nature. The leftmost interval
(including $A=0$ in the absence of a defect) corresponds to a typically large(r)
growth rate. Here, the instability derives from real eigenvalue
pairs. Connecting with Fig.~\ref{lc} and the case of
$A=0$, we recognize that this unstable mode is associated with
the breakup to vortex quadrupoles. As $A$ becomes increasingly
more negative to the left of the figure, other modes may, in turn,
dominate the instability dynamics (the ``bend'' in the stability
diagram represents such a ``take-over'' of the dominant instability
by a different mode; cf.~Fig.~\ref{lc}). However, it is observed 
that as $A$ increases on the positive side, the unstable real pair(s)
decrease in their real part and eventually cross through the origin
of the spectral plane, becoming imaginary and hence stabilizing the
RDS state. This is, once again, a key finding of our work, representing
the RDS stabilization.
However, as the (formerly unstable) eigenvalues bear a so-called  
`negative energy', upon climbing up the imaginary
axis, they may collide with eigenvalues associated with
`positive energy' modes (see, e.g., the discussion in pp.~56--58 of
Ref.~\cite{book2}).
This type of collision gives rise
to a complex eigenvalue quartet and a different (weak) oscillatory
dynamical instability, or a so-called Hamiltonian Hopf bifurcation;
see, e.g., the discussion of Ref.~\cite{goodman}. The latter scenario
leads to small instability bubbles, as the quartet may form, but
subsequently the eigenvalues may return to the imaginary axis,
splitting anew into two imaginary pairs.

\begin{figure}[tb] 
\begin{center}
\includegraphics[width=8.5cm]{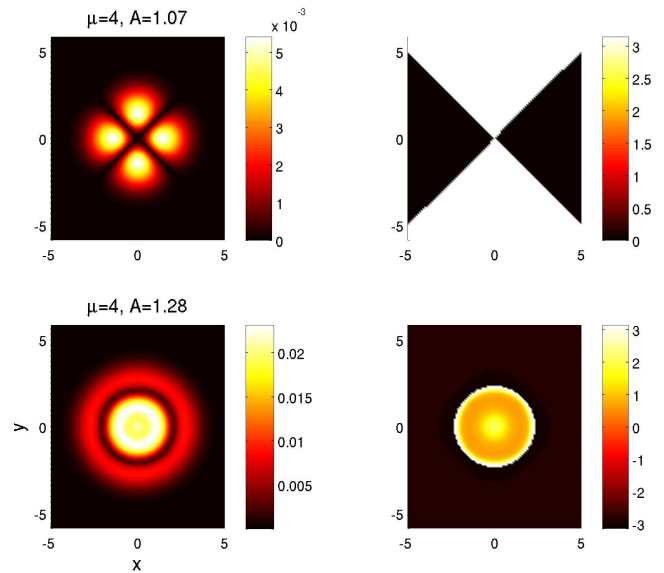}
\caption{(Color online) The most unstable modes of Fig.~\ref{css1}.
Top and bottom row of panels correspond, respectively, to the absolute
value (left subpanels) and phase (right subpanels) of the solutions
for the left and right cases depicted in Fig.~\ref{css1}.
}
\label{css2}
\end{center}
\end{figure}

\begin{figure}[tb] 
\begin{center}
\includegraphics[width=8.5cm]{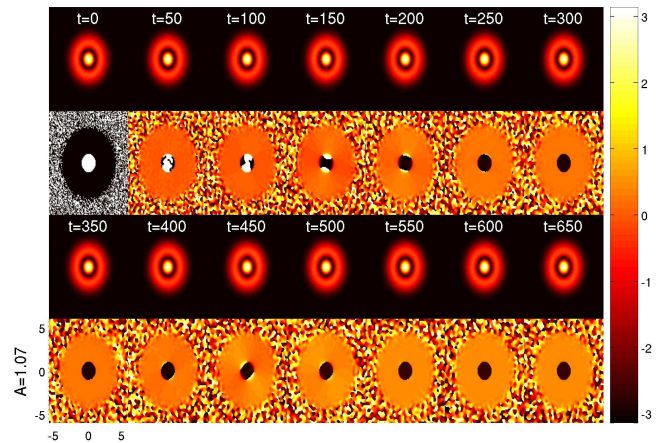}
\caption{(Color online) Dynamics of the state in the top panels of Fig.~\ref{css2}.
The odd panels depict the absolute value of the field while the even panels depict its phase.
The state is oscillating between the vortex quadrupole and the RDS, but very weakly.
}
\label{RK4A107}
\end{center}
\end{figure}

\begin{figure}[tb] 
\begin{center}
\includegraphics[width=8.5cm]{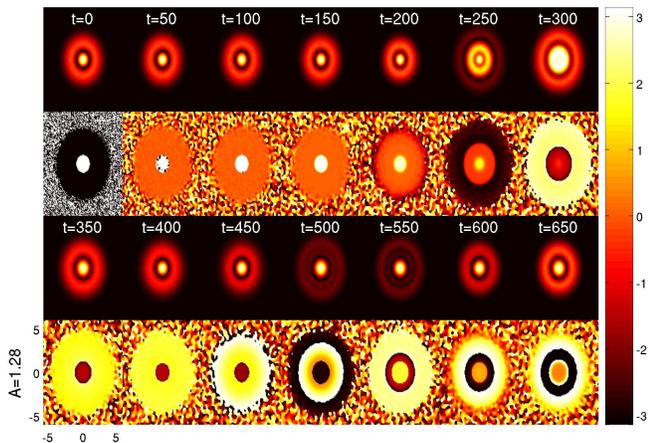}
\caption{(Color online) The same plot as Fig.~\ref{RK4A107} but for the state in the
bottom panels of Fig.~\ref{css2}. This state has a different nature of instability
from the one in Fig.~\ref{RK4A107}. The instability is like a vibrational mode.
}
\label{RK4A128}
\end{center}
\end{figure}

The two (exponential and oscillatory) instability scenarios
are illustrated in the two panels of Fig.~\ref{css1} for smaller
and larger values of $A$, respectively.
The most unstable mode of each state is shown in Fig.~\ref{css2},
illustrating the distinct nature of the instability in the
different scenarios.
The state at $A=1.07$ is in the same branch of $A=-1,0$ and 1, and
its instability leads to a deformation towards a vortex quadrupole state in a way similar as
the first plot of Fig.~\ref{evecs}. On the other hand,
the state at $A=1.28$ appears to have a different type of instability
that instead resembles a vibrational mode (the type of mode that
could be captured through a ring particle model). The time dynamics of the two
states are shown in Fig.~\ref{RK4A107} and Fig.~\ref{RK4A128} respectively.
In the former case, we observe the recurrent formation of a
vortex quadrupole (this is not immediately discernible in the
density but distinctly visible in the phase pattern),
in accordance with the identified unstable
mode. In the latter, indeed unstable vibrational dynamical characteristics
can be seen in the motion of the ring, which, however, appears to
maintain its radial structure.

\begin{figure}[tb] 
\begin{center}
\includegraphics[width=8.5cm]{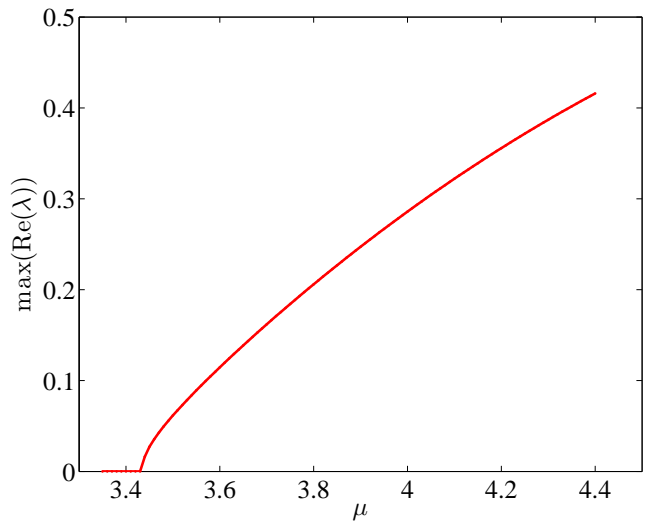}
\caption{(Color online) Cross section of max(Re($\lambda$)) at $A=0.5$. The solution starts to exist around $\mu=3.35$. Note that $A$ delays the set in of instabilities as $\mu$ is increased.
}
\label{LCA}
\end{center}
\end{figure}

A different cross section of the stability plane of Fig.~\ref{surface3}
is given in Fig.~\ref{LCA}, now for the case of $A=0.5$,
and varying the chemical potential $\mu$.
From this perspective, we 
observe that $A$ delays the onset of instabilities as $\mu$ is increased.
Another way to look
at the effects of $A$ and $\mu$
is that $A$ plays effectively the opposite role to that of $\mu$:
the increase of $A$ (for fixed $\mu$) drives the eigenmodes away from the
real axis and into the imaginary axis while the increase of the
chemical potential for fixed $A$ drives the eigenmodes away from the imaginary axis
and into the real axis, causing instability. We believe that
this discussion provides a unified perspective on the sources
of destabilization and the potential for re-stabilization of
the RDS.

\begin{figure}[tb] 
\begin{center}
\includegraphics[width=8.5cm]{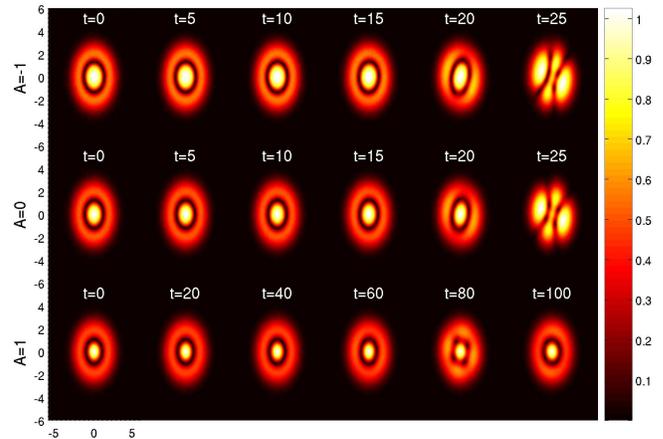}
\caption{(Color online) Time evolution of states at $\mu=4$ with $A=-1$, 0 and 1
(top to bottom rows of panels).
Note that the state of $A=1$ is significantly less unstable than those of $A=-1$ and
$A=0$, which have roughly the same instability growth rate. Note also that
all three states deform toward the vortex quadrupole state initially, although
the third one maintains an oscillatory pattern between a recurring ring
and a vortex quadrupole.
}
\label{RK4}
\end{center}
\end{figure}

\begin{figure}[tb] 
\begin{center}
\includegraphics[width=8.5cm]{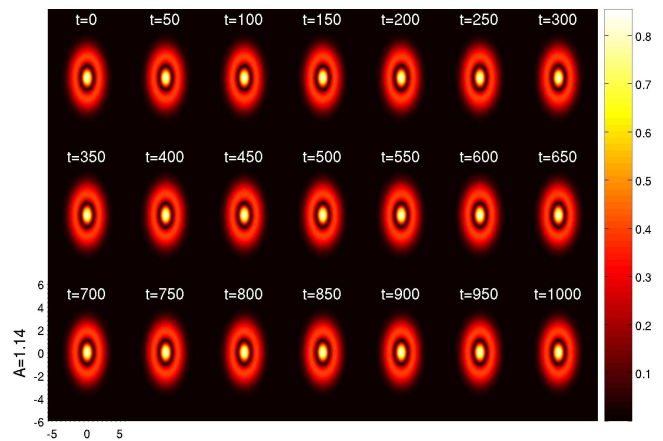}
\caption{(Color online) Time evolution of states at $\mu=4$ with $A=1.14$ which is in a
completely stable parametric interval. The state is shown to be stable up
to t=1000,  in agreement with our spectral results.
}
\label{RK4s}
\end{center}
\end{figure}

In all the cases considered, the stability conclusions were
also found to be consonant with the corresponding dynamics,
of which we now present a few additional case examples. In particular,
we study the dynamical evolution of states at $\mu=4$ for different values
of $A=-1,0,1$ (see Fig.~\ref{RK4})
and $1.14$ (see Fig.~\ref{RK4s}) to probe the effects of the variation of
$A$. Note that the cases of $A=-1$ and $0$ are about equally unstable at $\mu=4$ with $A=-1$
bearing a slightly larger growth rate. The case of $A=1$,
however, is very close to, albeit not within the stabilization regime.
On the other hand, the case of $A=1.14$ is fully stabilized.
We add a random perturbation to the states, ensuring that the number of atoms
in each case is, upon perturbation, 1.0013 times of the unperturbed one.
The results of the dynamical evolution of the former three cases are shown in Fig.~\ref{RK4}.
Note that both states for $A=-1$ and $A=0$ are relatively
quickly deformed around $t=25$ while the state for $A=1$ deforms only much
later around $t=70$, due to its weaker growth rate. In all three cases,
the states evolve initially towards the vortex quadrupole waveform.
While the former two states will quickly deform afterwards
and lose their radial structure, the third state can oscillate between the RDS state and the vortex quadrupole state for a much
longer time at least up to $t=1000$. A dynamical evolution of states at $A=1.14$, which is in a
completely stable parametric interval, is shown in Fig.~\ref{RK4s}.
The state is shown to be stable at least up to $t=1000$,
in agreement with the spectral findings and corroborating the
full stabilization achieved by the presence of the Gaussian
repulsive impurity.

\begin{figure}[tb] 
\begin{center}
\includegraphics[width=7.5cm]{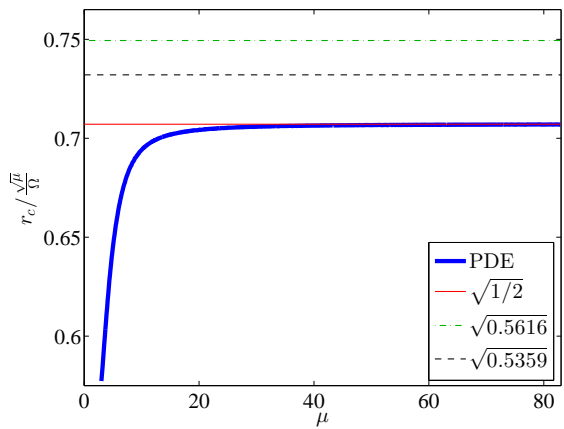}
\caption{(Color online) The location of the RDS scaled by $\sqrt{\mu}/\Omega$ as a function
of $\mu$ (thick solid blue line).
Note that the numerical values reach a limiting value of $1/\sqrt{2}$
(thin horizontal solid red line)
when $\mu$ is large. The particle picture can approximately describe the
$\sqrt{\mu}$ behavior and over estimates $r_c$, but nevertheless is still
an interesting approximate description of the RDS.
The particle approach using the perturbed Lagrangian method
[see Eqs.~(\ref{a}) and (\ref{b})] corresponds to the thin dotted-dash
green line while the solvability condition for the steady
state problem method [see Eq.~(\ref{neweq3})] is depicted by the
thin dashed black line.
}
\label{rc1}
\end{center}
\end{figure}

\begin{figure}[tb] 
\begin{center}
\includegraphics[width=7.5cm]{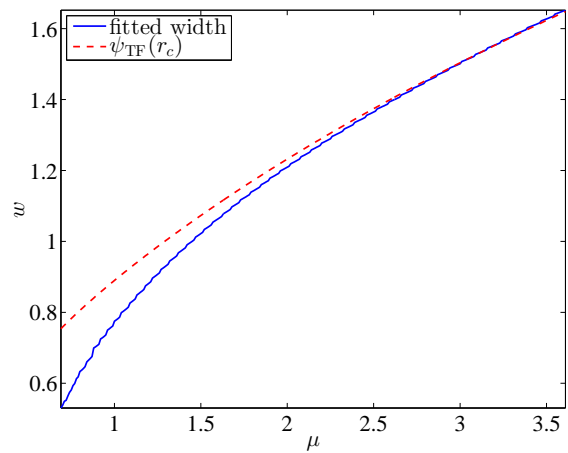}
\caption{(Color online) Dark soliton width $w$ as a function of the chemical potential $\mu$.
The blue solid line corresponds to fitting a profile
$\psi_{\rm TF}(r)\times\tanh(\sqrt{w}(r-r_c))$
to the PDE steady state for $\Omega=0.2$.
The red dashed line corresponds to the approximate value
of the background at the
location of the RDS, see Eq.~(\ref{eq:mu0}).
}
\label{fig:width}
\end{center}
\end{figure}

\subsection{The particle picture of the ring dark soliton}
\label{sec:results:particle}
We first study how the equilibrium location $r_c$ of the RDS changes with
chemical potential $\mu$, especially in the large density limit without
the perturbation potential, and compare the numerical results and the
particle picture predictions. Numerical results (for $\Omega=1$) suggest that
$r_c=\sqrt{\mu/2}$ (see thin horizontal red line)
in the large $\mu$ limit as shown in Fig.~\ref{rc1}.
As mentioned in Sec.~\ref{sec:sub:part}, a systematic analysis of
Eqs.~(\ref{a}) and (\ref{b}) yields 
the estimate
$r_c=\sqrt{0.5616\mu}/\Omega$ 
(see horizontal thin dotted-dash green line).
On the other hand, using the solvability condition for the steady
state problem described in Sec.~\ref{sec:sub:part}, one obtains the better prediction
of the RDS position $r_c=
\sqrt{0.5359\mu}/\Omega$; 
see Eq.~(\ref{neweq3})
and thin horizontal dashed black line in Fig.~\ref{rc1}.
It is important to mention that, although the above two particle approaches
are able to capture the $\sqrt{\mu}/\Omega$ behavior of $r_c$, they
do not lead to the precise numerical prefactor. This may be attributed to
the choice of the ansatz (\ref{pe}), where the width of the stationary
dark soliton is chosen to be $\sqrt{\mu}$. This selection corresponds
to the width of a dark soliton in a homogeneous background of density $\mu$.
However, due to the non-homogeneity of the BEC background, the RDS
placed at $r_c$ experiences a background density $\mu_0$ which can be
approximated using the TF 
regime (valid for large $\mu$) to be
\begin{equation}
\mu_0\approx\psi_{\rm TF}^2(r_c)=\mu-V(r_c)=\mu-\frac{1}{2}\Omega^2r_c^2.
\label{eq:mu0}
\end{equation}
For instance, in Fig.~\ref{fig:width} we show an example where we
extracted the width of the dark soliton for $\Omega=0.2$ as a
function of $\mu$. As it is clear from the figure, as $\mu$
increases, the width of the dark soliton converges to
$\sqrt{\mu_0}$ as prescribed in Eq.~(\ref{eq:mu0}).
Lastly, it is relevant to point out that, remarkably, the
adiabatic invariant theory of Ref.~\cite{kamch} properly
captures the asymptotic growth of the radius
of the RDS as $\mu$ increases. It is for that reason that
we will hereafter utilize the particle picture of Eq.~(\ref{PPextra})
and Ref.~\cite{kamch} for our further static and dynamics
considerations.

\begin{figure}[tb] 
\begin{center}
\includegraphics[width=8.5cm]{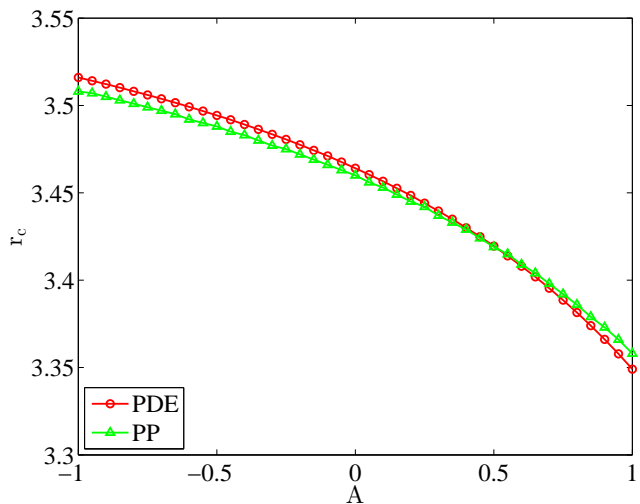}
\caption{
Position of the RDS as a function of $A$ for $\mu=24$.
The (red) circles correspond to the full PDE dynamics and
the (green) triangles to the particle picture (PP)
described in Sec.~\ref{sec:sub:part}.}
\label{rc2}
\end{center}
\end{figure}

We now study the effect of $A$ on $r_c$.
Figure~\ref{rc2} depicts $r_c$ as a function of $A$ for $\mu=24$.
It is clear that the particle picture can capture
the effect of $A$ fairly accurately. It is also observed that the
critical radius decreases in comparison to the $A=0$ limit, in the
presence of a repulsive defect, while the opposite is true in the
case of an attractive defect.

\begin{figure}[floatfix] 
\begin{center}
\includegraphics[width=8.5cm]{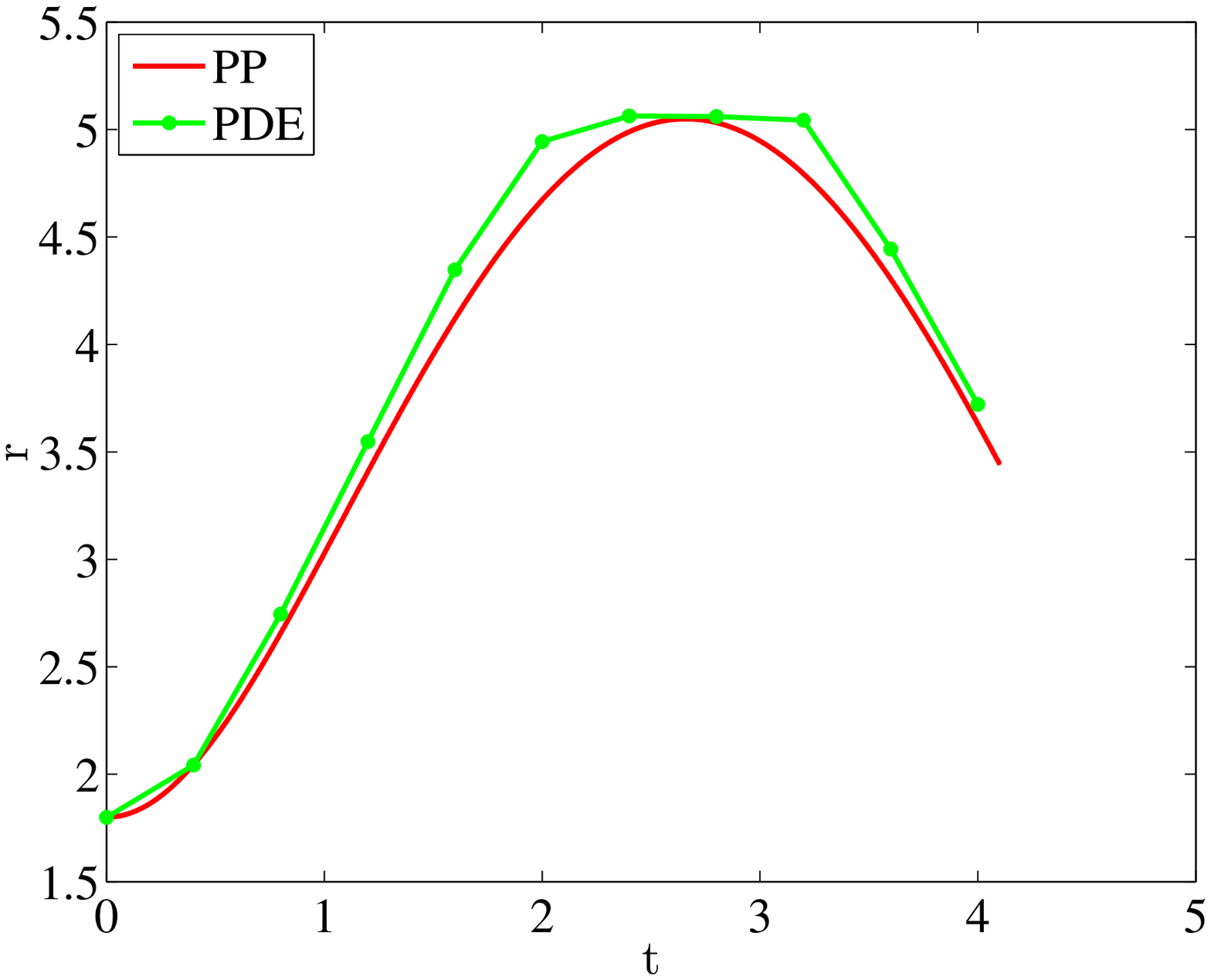}
\includegraphics[width=8.5cm]{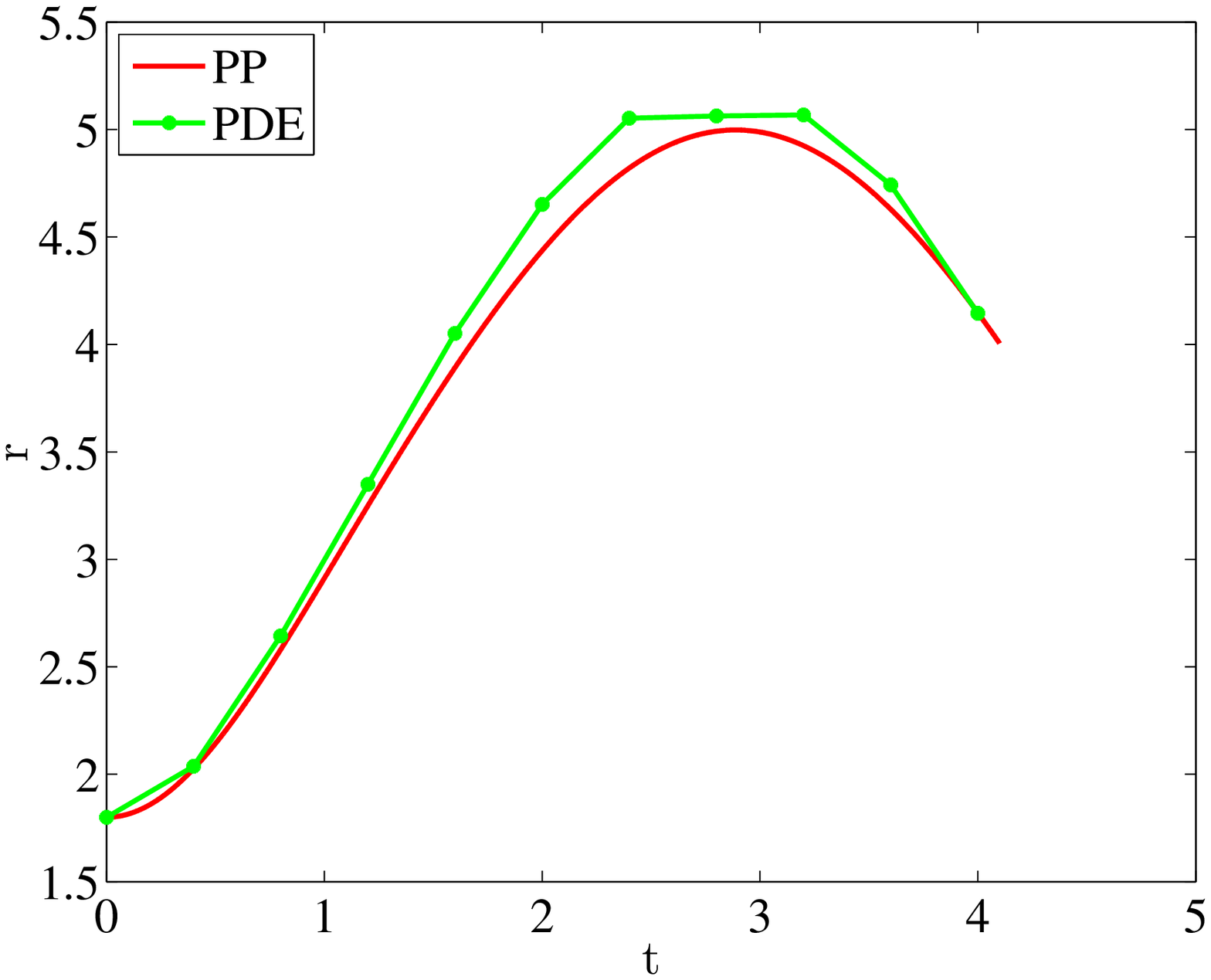}
\caption{Radial oscillatory motion of the RDS with $\mu=24$ for $A=0$
and $A=1$. The central radius of the RDS is extracted from the PDE
dynamics (green dots) and compared to the ODE evolution of the particle
picture (PP, red line) according to Eq.~(\ref{PPextra}).}
\label{rctime1}
\end{center}
\end{figure}

Finally, we study the radial oscillatory motion of the RDS in both the
case bearing and in that without the perturbation potential. We initialize
our  displaced RDS state by superposing a suitable hyperbolic
tangent profile to (i.e., multiplying
it with) the numerically exact ground state at the same
chemical potential $\mu=24$. Note that the RDS is unstable at such a high
chemical potential, therefore, we can only simulate the PDE dynamics for a
limited amount of time, before an instability leading to a
polygonal cluster of vortices ensues. The comparison of the PDE and the
particle picture dynamics for the cases of $A=0$ and $A=1$ are shown,
respectively, in the top and bottom panels of Fig.~\ref{rctime1}.
We see that the particle picture is able to capture
the essential PDE radial oscillation dynamics both with and
without the Gaussian barrier.

\section{Conclusions and Future Challenges}
\label{conclusion}

In this work, we studied the existence and stability of ring dark soliton
states, initially in the absence and subsequently
in the presence of a radially localized Gaussian perturbation potential.
We have systematically shown, via a combination of spectral analysis
and direct numerical simulations, that the ring dark soliton
can be stabilized by adding the perturbation potential with a suitable strength,
for all values of the chemical potential that we have considered herein.
Our systematic spectral analysis has also revealed why this stabilization
mechanism can only be effective near the linear limit of the system. It
has also revealed the potential for secondary instabilities (due to
pair collisions on the imaginary axis and complex eigenvalue quartets
emerging from them) due to the excited state nature of the ring.

An additional effort, significantly motivated by the potential
of the above method to lead to stable RDS vibrations,
was that of deriving dynamical
equations for their motion. We evaluated different techniques
to this effect, showcasing the fact that although all approaches
gave fairly similar results, the adiabatic invariant method
of Ref.~\cite{kamch} presented a distinct advantage in capturing
the radius of stationary rings.
A self-consistent perturbative technique (based on earlier work in
reaction-diffusion systems) was also adopted to that effect and was shown
to give reasonably accurate results in its comparison with the full
numerical
results.
Going beyond the ``stationary particle'' approach, allowing motion 
along the radial direction, an intriguing
goal for the future
may be to examine the ring soliton as a filamentary pattern embedded
in 2D, which, in addition to radial internal modes, may possess
bending ones (but without breaking). Such studies may in turn
enable the observation of collisions and deformations of rings
upon interactions, a topic that has been of interest also in
nonlinear optics~\cite{ektoras}.

Finally, it may well be relevant to explore settings beyond
the realm of two spatial dimensions, extending the present considerations
to the case of 3D solitonic or vortex rings and
other such patterns. Earlier work established how to construct
such states in isotropic and anisotropic 3D limits starting from
linear eigenstates~\cite{craso2}.
It is then of particular interest to continue
such states in the nonlinear realm and explore their spectral and
dynamical stability using tools similar to the ones proposed herein.
Efforts along these directions are currently in progress and will be
presented in future publications.

\begin{acknowledgments}

W.W.~acknowledges support from NSF (grant No. DMR-1208046).
P.G.K.~gratefully acknowledges the support of
NSF-DMS-1312856, as well as from
the US-AFOSR under grant FA9550-12-1-0332,
and the ERC under FP7, Marie Curie Actions, People,
International Research Staff Exchange Scheme (IRSES-605096).
P.G.K.'s work at Los Alamos is supported in part by the U.S. Department of Energy.
R.C.G.~gratefully acknowledges the support of NSF-DMS-1309035.
The work of D.J.F.~was partially supported by the Special Account
for Research Grants of the University of Athens.
The work of T.J.K.~was supported in part by NSF
grant DMS-1109587.
M.M.~gratefully acknowledges support from the provincial Natural
Science Foundation of Zhejiang (LY15A010017) and the National
Natural Science Foundation of China (No.~11271342).

\end{acknowledgments}

\end{document}